\documentclass[conference]{IEEEtran}
\IEEEoverridecommandlockouts
\usepackage{cite}
\usepackage{amsmath,amssymb,amsfonts}
\usepackage{algorithmic}
\usepackage{graphicx}
\usepackage{textcomp}
\usepackage{xcolor}
\usepackage{tikz}
\usetikzlibrary{shapes}
\usepackage{pgfplots}
\usepackage{comment}
\def\BibTeX{{\rm B\kern-.05em{\sc i\kern-.025em b}\kern-.08em
    T\kern-.1667em\lower.7ex\hbox{E}\kern-.125emX}}

\newcommand{\expjwT}[1]{e^{-j\omega T #1}}
\newcommand{\exppjwT}[1]{e^{j\omega T #1}}

\newcommand{\intovertwopi}[1]{\int_{-\pi}^{\pi} #1 d(\omega T)}

\tikzstyle{startstop} = [rectangle, rounded corners, minimum width=3cm, minimum height=0.5cm,text centered, draw=black, font=\footnotesize]
\tikzstyle{io} = [trapezium, trapezium left angle=70, trapezium right angle=110, minimum width=3cm, minimum height=1cm, text centered, draw=black, , font=\footnotesize]
\tikzstyle{process} = [rectangle, minimum width=2cm, minimum height=0.5cm, text centered, draw=black, font=\footnotesize]
\tikzstyle{decision} = [diamond, minimum width=2.5cm, minimum height=1.5cm, inner sep=0, text centered, draw=black, font=\footnotesize]
\tikzstyle{arrow} = [->,>=latex]

\definecolor{superlightgray}{rgb}{0.96, 0.96, 0.96}
\definecolor{verylightgray}{rgb}{0.9, 0.9, 0.9}
\definecolor{applegreen}{rgb}{0.55, 0.71, 0.0}

\begin{document}

\title{Efficient Design and Implementation of Fast-Convolution-Based Variable-Bandwidth Filters}

\author{\IEEEauthorblockN{Oksana Moryakova and H\aa kan Johansson}
\IEEEauthorblockA{Division of Communication Systems, Department of Electrical Engineering, Link\"{o}ping University,
Link\"{o}ping, Sweden\\
Emails: \{oksana.moryakova, hakan.johansson\}@liu.se
}
}

\maketitle

\begin{abstract}
This paper introduces an efficient design approach for a fast-convolution-based variable-bandwidth (VBW) filter. The proposed approach is based on a hybrid of frequency sampling and optimization (HFSO), that offers significant computational complexity reduction compared to existing solutions for a given performance. The paper provides a design procedure based on minimax optimization to obtain the minimum complexity of the overall filter. A design example includes a comparison of the proposed design-based VBW filter and time-domain designed VBW filters implemented in the time domain and in the frequency domain. It is shown that not only the implementation complexity can be reduced but also the design complexity by excluding any computations when the bandwidth of the filter is adjusted. Moreover, memory requirements are also decreased compared to the existing frequency-domain implementations. 
\end{abstract}

\begin{IEEEkeywords}
Variable bandwidth filter, fast convolution, frequency-domain design, time-varying systems, overlap-save, multirate filter banks
\end{IEEEkeywords}

\section{Introduction} \label{sec:introduction}
Due to the increasing demand for reconfigurable systems in the contemporary world of technologies, variable digital filters (VDFs) are required in many digital signal processing applications, for example, in medical devices \cite{Indrakanti_2018} and communication systems \cite{RAGHU19, Renfors_2014}. The main advantage of these filters over regular digital filters is that they offer variability of the frequency response by adjusting only one or a few parameters without online filter design. Most recent papers focused on VDFs \cite{Indrakanti_2018, RAGHU19, Renfors_2014} have shown that this approach allows to significantly reduce implementation complexity and thereby hardware complexity compared to regular filters requiring online design for every new specification. Nevertheless, implementations of VDFs in the time domain may still cause a rather high computational complexity for stringent requirements.

Research works on finite-impulse-response (FIR) filters have shown that a filter can be implemented in the frequency domain using the fast convolution (FC) with much lower complexity than in the time domain \cite{Shynk_1992, Ishihara_2011, Kovalev_2017, Johansson_23}. The most widely used techniques are the overlap-save (OLS) and overlap-add (OLA) methods which employ the discrete Fourier transform (DFT) and its inverse (IDFT) \cite{Oppenheim_DTSP, Daher_2010}. The FC-based VDFs \cite{Renfors_2014, Moryakova_2023} have shown significant reduction of the computational complexity compared to time-domain implementations. 

Typically, regardless of the implementation, VDFs are designed 
by optimizing the impulse response values to satisfy specification requirements for the frequency response, hereafter this is also referred as \textit{time-domain design}. In this case, the filter DFT coefficients are given by the DFT of the optimized impulse response.
For frequency-domain implementations, this fact leads to less efficient implementation and update of variable coefficients in terms of computational complexity. Thus, this work is devoted to \textit{frequency-domain design} of FC-based VDFs, specifically variable-bandwidth (VBW) filters, by instead optimizing the filter DFT coefficients. The OLS method is considered here due to somewhat less implementation complexity compared to the OLA method \cite{Johansson_23}.

The main contributions of the paper are as follows.

\begin{enumerate}
	\item An approach to FC-based VBW filter design allowing to significantly reduce both implementation and design complexity is proposed. The core idea of the proposal is to design a filter based on a hybrid of frequency sampling and optimization (HFSO) with the aim of utilizing the same set of transition band values for different bandwidths in the implementation. The proposed HFSO method consists of designing a VBW filter by optimizing only one set of the transition band samples while the passband and stopband values are directly obtained by sampling the desired frequency response. This means that the magnitude of the filter DFT coefficients for various bandwidths are sets of the same values, i.e., ones, zeros, and the optimized transition band values. This approach leads to a very simple update of the filter DFT coefficients when the bandwidth is varied, which was the drawback of the variable-weight-based structure with the lowest implementation complexity proposed in \cite{Moryakova_2023} requiring computations for every new value of the variable parameter. Hence, our proposed approach allows to simultaneously lower the implementation complexity, by cutting the number of general multiplications per sample, and significantly reduce the design complexity when the bandwidth is adjusted. Moreover, the proposed approach allows to eliminate the dependence of the implementation and design complexities on the variable bandwidth range while it is one of the main limitations in the existing VBW filter implementations \cite{Lowenborg_2006, Moryakova_2023}.
	
	\item A systematic procedure for FC-based VBW filter design using the HFSO approach is provided. 
	A similar approach based on the frequency sampling has been utilized for regular FIR filter design \cite{Rabiner_1970, Harris_1998, SalcedoSanz_2007, Belorutsky_2016}. However, the authors did not consider FIR filter implementations using the OLS or OLA methods, which together with the frequency sampling method make the filter a linear periodic time-varying (LPTV) system. Design of these systems has to include measuring the distortion and aliasing functions, which can be derived from a multirate filter bank representation, or a set of time-invariant impulse responses of the corresponding periodically time-varying impulse-response (PTVIR) representation. Although it was mentioned in \cite{Renfors_2014} that a frequency-domain designed FC-based filter, causing cyclic distortions, 
	can be modeled as an LPTV system,
	the authors did not provide details whether the corresponding time-invariant responses have been controlled during the design procedure. In our proposed approach, the overall filter is designed by ensuring that each of the corresponding time-invariant responses meets the specification using minimax optimization.
\end{enumerate}

The rest of the paper is organized as follows. Section \ref{sec:OLS} gives a brief overview of the overlap-save method while Section \ref{sec:prop_approach} presents the proposed approach of an FC-based VBW filter design using the HFSO method. Section \ref{sec:examples} provides an example and computational-complexity analysis. Finally, Section \ref{sec:conclusion} concludes the paper.

\section{Overlap-Save Method}\label{sec:OLS}
The main concept of the FC using the OLS method is that the input signal  is divided into overlapping segments $x_m(n)=x(n+mM)$, $n=0,1,..., N-1$, where $m$ is a segment index and the overlapping part is $N-M$. Then, for each segment, the following computations are carried out.
\begin{enumerate}
	\item The segment $x_m$ is transformed via an $N$-point DFT.
	\item The DFT coefficients $X_m(k)$ are multiplied by the filter DFT coefficients $H(k)$.
	\item An $N$-point IDFT is performed.
	\item The first $N-M$ samples of the resulting block are discarded, so that the output segments of length $M$ are no longer overlapping. 
	\item The output sequence is obtained by concatenating the resulting segments as $y(n)=\sum_{m=0}^{\infty}y_m(n-mM)$.
\end{enumerate}

In classical OLS filtering, the system is intended to be time-invariant, i.e., the aliasing error is zero. In this case, the  DFT coefficients $H(k)$ of an FIR filter correspond to the $N$-point DFT of the impulse response $h(n)$ of length $L$, designed by any of the available methods \cite{Oppenheim_DTSP}. This means $h(n)=0$ for $n=L,...,N-1$, where $L=N-M+1$.

In this paper, the proposed HFSO design approach\footnote{The HFSO design approach will be discussed in details in Section \ref{subsec:design_FS}.} restricts the frequency response to take on fixed values for certain frequencies. Even if the effective order of the underlying filter is $L-1$, after an $N$-point IDFT, this results in an $N$-length impulse response (although the impulse response values are quite small for $L\leq n \leq N-1$) and thereby makes the overall system an $M$-periodic time-varying system, i.e., the filter impulse response coefficients change from sample to sample \cite{Daher_2010}. In these systems, aliasing cannot be cancelled but can be suppressed to any desired level through a proper design, which is considered in this paper. An LPTV system can be represented using a PTVIR representation
corresponding to a set of time-invariant impulse responses $h_n(q)=h_{n+M}(q)$, $n = 0, 1,..., M-1$, $q = 0, 1, ..., N-1$ (or their corresponding frequency responses $H_n(\exppjwT{})$) \cite{Daher_2010, Vaidyanathan_MSFB, Johansson_23}. This representation can be a better indicator of the worst-case time-domain error of the overall system comparing to the multirate filter-bank representation in terms of distortion and aliasing \cite{Johansson_23}. 

\begin{figure}
	\centering
	\includegraphics[width=\linewidth]{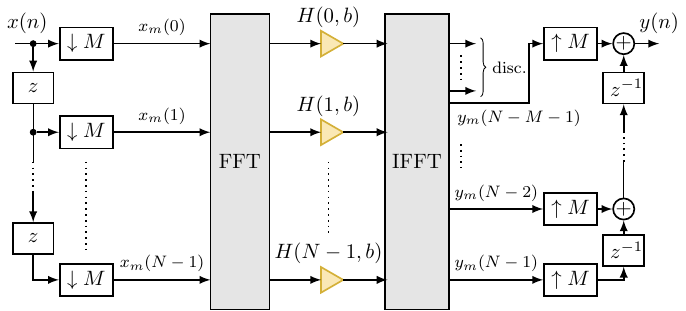}
	\caption{Frequency-domain implementation of a VBW filter using the overlap-save method \cite{Moryakova_2023}.}
	\vspace{-0.1cm}
	\label{fig:VBW_FD_original}
\end{figure}

\begin{figure}
	\centering
	\includegraphics[width=\linewidth]{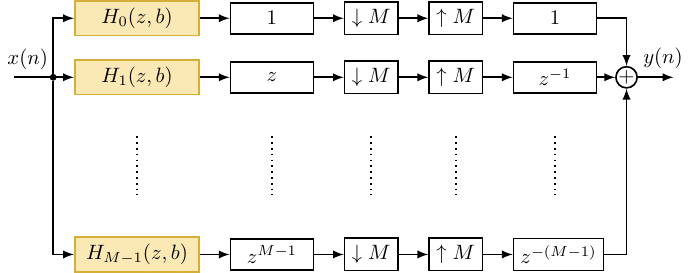}
	\caption{PTVIR representation of the scheme in Fig. \ref{fig:VBW_FD_original}.}
	\vspace{-0.25cm}
	\label{fig:VBW_FD_PTVIR}
\end{figure}

\section{Proposed Approach to Design a VBW Filter Implemented Using the OLS method}
\label{sec:prop_approach}

\subsection{VBW Filter Implemented Using the OLS method}
\label{subsec:VBW_filt}
The desired frequency response of a VBW filter of effective length  $L$ is considered here as
\begin{equation}
	D(\exppjwT{}, b) = 
	\begin{cases}
		\expjwT{(L-1)/2},  & \omega T \in[0, b-\Delta/2],\\
		0,  & \omega T \in[b+\Delta/2, \pi],
	\end{cases}
	\label{eq:VBW_des}
\end{equation}
where $b\in[b_l, b_u]$ is the center of the transition band, whereas  $\omega_cT = b-\Delta/2$ and $\omega_sT = b+\Delta/2$ represent variable passband and stopband edges, respectively. The transition width $\Delta$ is assumed to be fixed.

The VBW filter frequency-domain implementation proposed in \cite{Moryakova_2023} is shown in Fig. \ref{fig:VBW_FD_original}. This structure utilizes $N$ variable DFT coefficients $H(k,b)$ that have to be recomputed whenever $b$ is changed. As was mentioned in Section \ref{sec:OLS}, this implementation can be equivalently represented using the PTVIR representation, which is shown in Fig. \ref{fig:VBW_FD_PTVIR}. The output of the system can be written as
\begin{equation}
	y(n, b)=\frac{1}{2\pi}\intovertwopi{H_n(\exppjwT{}, b)X(\exppjwT{})\exppjwT{n}},
\end{equation}
where $H_n(\exppjwT{}, b)$ are modified responses from \cite{Johansson_23} by introducing the variable parameter $b$. They can be expressed as
\begin{equation}
	H_n(\exppjwT{}, b)=\expjwT{n}\sum_{q=0}^{N-1}d_n(q, b)\expjwT{q}
	\label{eq:OLS_Hn}
\end{equation}
with $d_n(q,b)$ being the impulse responses of $z^n H_n(z, b)$ and circular versions of each other, i.e., $d_{n+m}(q, b)=d_n(\mod(q+m,M), b)$. This allows to avoid computations of $d_n(q,b)$ for every $n$ in the design. 
The last response $d_{M-1}(q,b)$ corresponds to the IDFT of $H(k,b)$, i.e.,
\begin{equation}
	d_{M-1}(q, b) = \frac{1}{N}\sum_{k=0}^{N-1}H(k, b)e^{j2\pi qk/N}.
	\label{eq:OLS_dn}
\end{equation}
The DFT coefficients $H(k,b)$ are obtained using the HFSO approach that will be described in the following subsection.

\subsection{Proposed HFSO Approach}
\label{subsec:design_FS}
\begin{figure}
	\centering
	\includegraphics[trim={5cm 7cm 5.3cm 6.5cm}, clip, width=\linewidth]{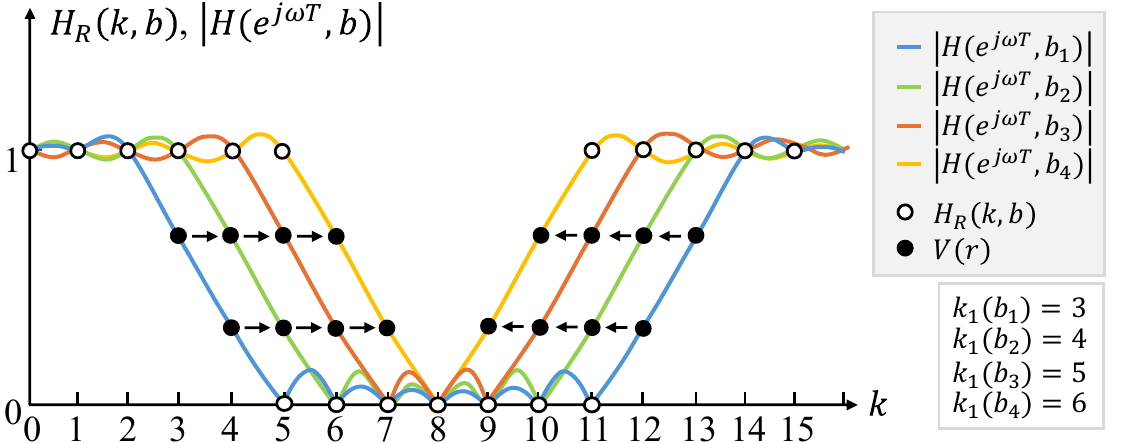}
	\caption{HFSO approach for a VBW filter.}
	\vspace{-0.35cm}
	\label{fig:FSbased_method}
\end{figure}

The proposed HFSO approach consists of two parts: sampling of the passband and stopband of the desired frequency response $D(\exppjwT{}, b)$ in \eqref{eq:VBW_des} and optimization of the samples belonging to the transition band, which is generally not specified in the desired response, in order to minimize the passband and stopband ripples of all responses $H_n(\exppjwT{}, b)$, $n=0,1,...,M-1$, for $b\in[b_l, b_u]$. Thus, the coefficients $H(k,b)$ are given as  
\begin{equation}
	H(k,b) = H_R(k,b)e^{-j\frac{2\pi k (L-1)/2}{N}},
	\label{eq:FS_Hk_exp}
\end{equation}
with the magnitude response samples $H_R(k,b)$ given by 
\begin{align}
	H_R(k,b) =
	\begin{cases}
		1,  & k\in k_{p}(b),\\
		V(k-k_1(b)),  & k\in k_t(b),\\
		0,  & k\in k_s(b),
	\end{cases}
	\label{eq:FS_Hk_real}
\end{align}
where $k_p(b) = [0,k_1(b)-1] \cup [N-k_1(b)+1,N-1]$, $k_t(b)=[k_1(b), k_2(b)] \cup [N-k_2(b), N-k_1(b)]$, and $k_s(b)=[k_2(b)+1, N-k_2(b)-1]$ are the passband, transition band, and the stopband regions, respectively. Here, $k_1(b)$ and $k_2(b)$ are the first and the last sample indices of the transition band, correspondingly, given by
\begin{align}
	k_1(b)=\frac{b-\Delta/2}{2\pi}N+1, \quad
	k_2(b)=\frac{b+\Delta/2}{2\pi}N-1,
	\label{eq:k_var}
\end{align}
where the values of $\Delta$ and $b$ are assumed to be discretized so that  $\Delta=\Delta_N \times2\pi/N$ and $b=b_N\times2\pi/N$ with $\Delta_N=k_2(b)-k_1(b)+2$ and $b_N\in[b_{N,l}, b_{N,u}]$ being the fixed transition width and variable parameter in terms of frequency bins, $\Delta_N/2\leq b_{N,l} < b_{N,u} \leq N/2-\Delta_N/2-1$. It is assumed that the transition band samples can be utilized for different values of $b$ as it is illustrated in Fig. \ref{fig:FSbased_method}. Therefore,  $V(k-k_1(b))$ do not depend on $b$ and can be written as $V(r)$, $r=0,\dots, L_V-1$ with $L_V=\Delta_N-1$. 

Considering that the VBW filter impulse response is real-valued, the responses $d_n(q,b)$ are also real-valued. Then, $d_{M-1}(q,b)$ in \eqref{eq:OLS_dn} is given by
\begin{align}
	&d_{M-1}(q,b) = \frac{1}{N}\Bigg[1 + 2\sum_{k=1}^{k_1(b)-1}\cos\Bigg(\frac{2\pi k}{N}\Bigg(q-\frac{L-1}{2}\Bigg)\Bigg) \notag\\
	&+ 2\sum_{k=k_1(b)}^{k_2(b)}V(k-k_1(b))\cos\Bigg(\frac{2\pi k}{N}\Bigg(q-\frac{L-1}{2}\Bigg)\Bigg)\Bigg].
\end{align}

\subsection{Minimax Design}
\label{subsec:design_minmax}
In order to design an LPTV system modeled as a set of time-invariant frequency responses, one needs to ensure that each of the responses approximates the desired response \cite{Johansson_23}. Therefore, for a VBW filter design, each $H_n(\exppjwT{}, b)$ has to approximate 
the desired frequency response $H_d(\exppjwT{}, b)$, given by\footnote{The desired response $H_d(\exppjwT{}, b)$ in \eqref{eq:OLS_Hnd} differs from $D(\exppjwT{}, b)$ in \eqref{eq:VBW_des} by an additional delay of $M-1$ samples due to the OLS implementation.}  
\begin{equation}
	H_{d}(\exppjwT{}, b) = 
	\begin{cases}
		\expjwT{(\frac{L-1}{2}+M-1)},  & \omega T\in [0, b-\Delta/2], \\
		0, \quad &  \omega T\in [b+\Delta/2, \pi].
	\end{cases}
	\label{eq:OLS_Hnd}
\end{equation}
In this paper, they are approximated in the minimax sense. 
This means that the overall filter is here designed by solving the following approximation problem. For given $L$ and $N$, find $L_V$ transition band values $V(r)$ of the DFT coefficients $H(k,b)$ as well as $\delta$ to
\begin{align}
	& \textbf{minimize } \delta\notag\\
	& \textbf{subject to } |E_n(\exppjwT{}, b)|\leq \delta
	\label{eq:minimax}
\end{align} 
for $\omega T\in[0, b-\Delta/2] \cup [b+\Delta/2, \pi]$, $b\in[b_l, b_u]$, and $n=0,1,...,M-1$, where $E_n(\exppjwT{}, b)$ is the error function, given by
\begin{equation}
	E_n(\exppjwT{}, b) =H_n(\exppjwT{}, b)-H_d(\exppjwT{},b)
	\label{eq:err_min}
\end{equation}
with $H_n(\exppjwT{}, b)$ as in \eqref{eq:OLS_Hn}, $H_d(\exppjwT{},b)$ as in \eqref{eq:OLS_Hnd}.
The filter will meet the specification if $\delta$ after the optimization satisfies $\delta\leq\delta_E$, where $\delta_E$ is the maximum prespecified approximation error.

\subsection{Implementation and Design Complexity}
\label{subsec:complexity}
\begin{table*}[t]
	\renewcommand{\arraystretch}{1.1}	
	\centering
	\caption{Implementation and Design Complexity Expressions for the VBW Filter Designed by the Proposed Technique}
	
	\begin{tabular}{c c c c}
		\hline 
		Fixed multiplication rate $R_{mf}$ & Variable multiplication rate $R_{mv}$ & Addition rate $R_a$ & Memory\\
		\hline \hline 
		$\frac{N\log_2(N)-(3/2)N+4}{N-L+1}$ & $\frac{2L_V}{N-L+1}$ & $\frac{3N\log_2(N)-(7/2)N+8}{N-L+1}$ & $L_V$ \\ 
		\hline
	\end{tabular}
	\vspace{-0.35cm}
	\label{tab:complexity_rates}
\end{table*}

The overall complexity of the frequency-domain implementations consists of DFT/IDFT transforms and complex multiplications by $H(k,b)$. In this paper, we consider that the former is implemented using the split-radix fast Fourier transform (FFT) algorithm, the complex multiplication inside the transformation is implemented using 3 real fixed multiplications and 3 real additions, the segments $x_m$ and $y_m$ are real-valued and $N=2^Q$. Therefore, each FFT and its inverse (IFFT) require $C_{mf,F}=(1/2)N\log_2(N)-(3/2)N+2$ multiplications and $C_{a,F}=(3/2)\log_2(N)-(5/2)N+4$ additions \cite{Sorensen_1987}. The multiplications by $H(k,b)$ can be implemented as two consecutive multiplications: a variable (general) multiplication by real $H_R(k,b)$ and a fixed complex multiplication by $e^{-j2\pi k (L-1)/(2N)}$. The former is variable and real, thus, it can be implemented using two real general multiplications. The latter is complex and fixed, therefore, it can be implemented using 3 real fixed multiplications and 3 real additions. Moreover, for particular combinations of $L$ and $N$ such that $(L-1)/N=1/(2c)$, $c=1,2,...,N/4$, the exponents can be implemented even cheaper (e.g., for $k=mc$, $m=0,1,...,N/c-1$, there are no computations required, and for $k=mc+\lfloor c/2 \rfloor$, $c\geq2$, only two real multiplications and two real additions can be used). Additionally, for zero-valued $H_R(k,b)$, there are no multiplications by the exponents needed. Thus, the number of fixed multiplications is varied depending on $b$ and the ratio $(L-1)/N$. Here, we specify the maximum possible number for fair comparison. Considering that the DFT coefficients $X_m(k)$ and $H(k,b)$ are conjugate symmetric and ones and zeros in $H_R(k,b)$ do not require any computations, the number of fixed multiplications, general multiplications and additions are $C_{mf,H}=3N/2$, $C_{mv,H}=2L_V$, and $C_{a,H}=3N/2$, respectively. Therefore, the total implementation complexity per sample is expressed as a fixed multiplication rate $R_{mf}$, variable multiplication rate $R_{mv}$, and addition rate $R_a$ summarized in Table \ref{tab:complexity_rates}. 

For the proposed approach, there are no computations when $b$ is altered. Therefore, the design complexity includes only memory to store $L_V$ values of $V(r)$.

\subsection{Design Procedure Minimizing the Overall Complexity}
\label{subsec:des_proc}
To minimize the overall complexity per sample, the values of $M$, $L$, and $N$ have to be determined. 
The value of $L$ is obtained based on the filter order $N_D$ as $L=N_D+1$, where the initial $N_D$ can be estimated by \cite{Bellanger_84}
\begin{equation}
	\widehat{N}_D = -\frac{4\pi \log_{10}(10\delta_p \delta_s)}{3\Delta}.
	\label{eq:filt_order_estim}
\end{equation}
The optimal value of $N$ is estimated using $\widehat{N}=0.9L\log_2(L)$ \cite{Johansson_23} and rounded to the nearest $2^Q$. The value of $M$ is computed as $M=N-L+1$.

Further, the approximation problem in \eqref{eq:minimax} is a convex optimization problem which guarantees that the solution is globally optimal in the minimax sense that can be solved using any regular solver for such problems. In this paper, the optimization problem is solved using the optimization toolbox in MATLAB with $\omega T$ discretized into $K$ grid points. 
After the optimization, if the approximation error $\delta<\delta_E$, one should
% store the obtained values of $V(r)$ and the corresponding $L$, $N$, $M$ values, 
reduce the order $N_D$ and repeat the optimization. If the approximation error $\delta>\delta_E$, the order $N_D$ must be increased, the values of $L$, $N$, $M$ recomputed and the optimization repeated until the $\delta\leq\delta_E$.

\section{Example}
\label{sec:examples}
\begin{figure}[t]
	\centerline{\includegraphics[trim={0cm 0cm 0cm 0.22cm}, clip,width=\linewidth]{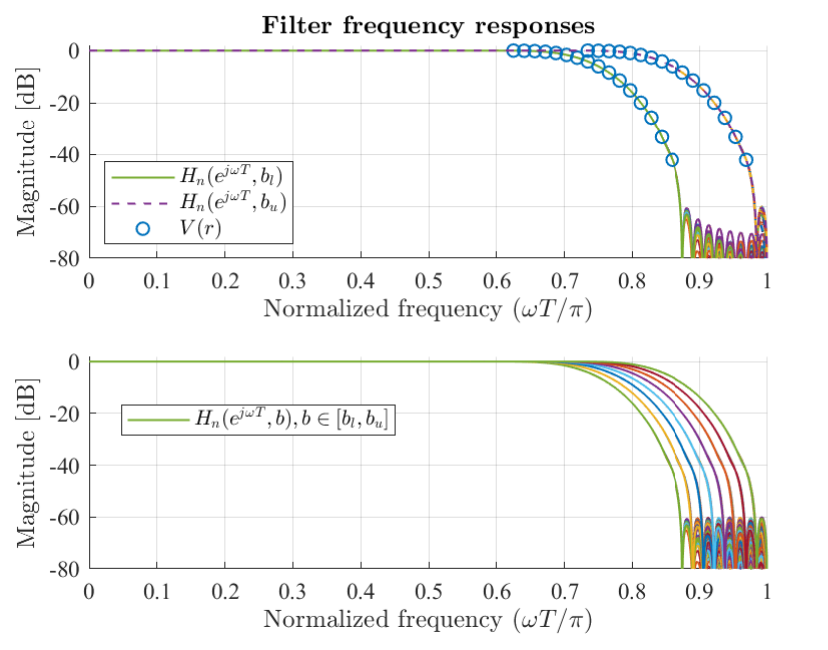}}
	\centerline{\includegraphics[trim={0cm 0cm 0cm 5.5cm}, clip,width=\linewidth]{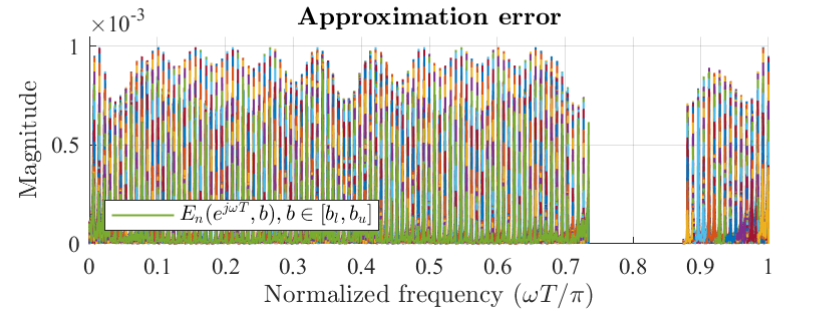}}
	%\vspace{-0.2cm}
	\caption{Modulus of the filter transition band samples $V(r)$, frequency responses $H_n(\exppjwT{}, b)$, and errors $E_n(\exppjwT{},b)$ in \eqref{eq:err_min} for $b\in[0.75\pi, 0.8594\pi]$ in the example.}
	\vspace{-0.3cm}
	\label{fig:narrow}
\end{figure}

In this example, we apply the proposed design for a VBW filter and compare the so obtained OLS implementation with time-domain designed VBW filters implemented using the Farrow structure \cite{Lowenborg_2006} and the OLS method \cite{Moryakova_2023}. 

The specification of the VBW filter is given as follows: $\Delta=0.25\pi$,  $\delta_p=\delta_s=\delta_E=0.001$, $b\in[0.75\pi, 0.8594\pi]$. According to the estimation expressions in Section \ref{subsec:des_proc}, the estimated filter order $\widehat{N}_D=28$, which gives $\widehat{N}=127$, and we choose $N=128$. Thus, the transition band and bandwidth in terms of bins are $\Delta_N=16$ and $b_N\in[b_{N,l}, b_{N,u}]=[48,55]$, respectively.
The passband and stopband regions are discretized to $K=1000$ grid points.

After the optimization outlined in Section \ref{subsec:des_proc}, $L=33$, $N=128$, $M=96$. Figure \ref{fig:narrow} plots the magnitude responses $H_n(\exppjwT{},b)$ of the VBW filter, where the top plot shows $M$ responses $H_n(\exppjwT{},b)$ for $b=b_l$ and $b=b_u$ (for better visibility) and corresponding transition band samples $V(r)$, the mid plot shows all $H_n(\exppjwT{},b)$ for $b\in[b_l, b_u]$, and the bottom plot shows the error $E_n(\exppjwT{},b)$ in \eqref{eq:err_min}. It is seen that the stopband level of all the responses is less than 60 dB. 

\begin{figure}[t]
	\centerline{\includegraphics[trim={0cm 0cm 0cm 0.22cm}, clip,width=\linewidth]{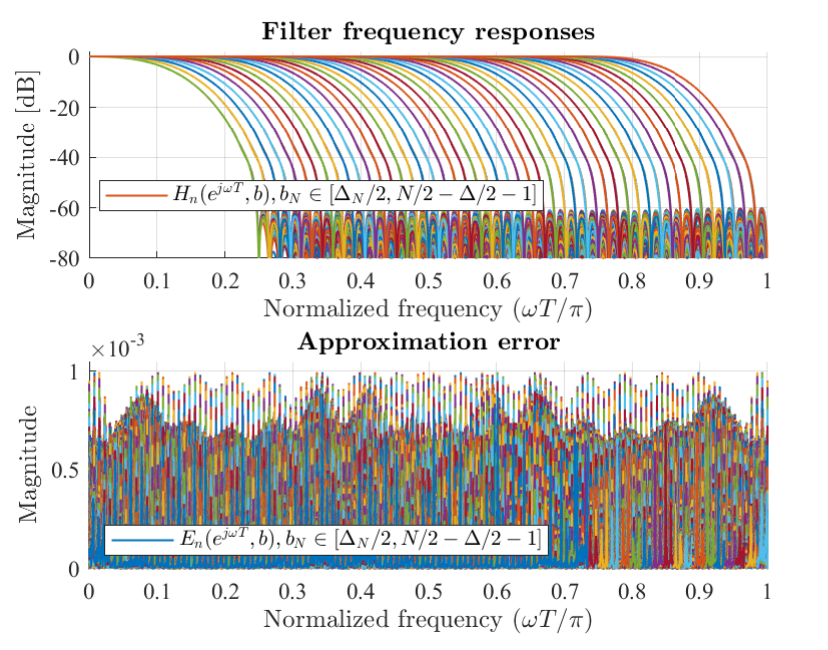}}
	\caption{Modulus of the frequency response and the error in \eqref{eq:err_min} for the entire variable band case in the example.}
	\label{fig:wide}
	\vspace{-0.cm}
\end{figure}

The implementation and design complexity rates are listed in Tables \ref{tab:example_imp} and  \ref{tab:example_des}, respectively. 
Since $(L-1)/N=1/4$ in this example, the exponents in \eqref{eq:FS_Hk_exp} can be implemented using 2 real multiplications and 2 additions only for every odd $k$ whereas there are no computations for every even $k$.
It is seen that the proposed approach shows significant savings in both complexities compared to other implementations. Moreover, the proposed approach allows to design and implement a VBW filter with the same complexity regardless the range of the variable parameter $b$, while the filters in \cite{Lowenborg_2006} and \cite{Moryakova_2023} cover only narrow range of variable parameter $b$ for a reasonable complexity. This means that for the entire variable band case, which is illustrated in Fig. \ref{fig:wide}, the complexity savings offered by the proposed method will be even much higher.

\begin{table}[t]
	\renewcommand{\arraystretch}{1.05}
	\caption{Implementation Complexity for the VBW Filter in the Example}
	\begin{center}
		\begin{tabular}{l c c c c c c c}
			\hline 
			Design/Impl. & $L$ & $N$ & $M$ & $P$ & $R_{mf}$ & $R_{mv}$ & $R_{a}$ \\
			\hline \hline
			TD/TD \cite{Lowenborg_2006} & 29 & - & - & 4 & 75 & 4 & 145 \\
			TD/FD \cite{Moryakova_2023} & 29 & 128 & 100 & - & 5.2 & 1.9 & 23.8 \\	
			\hline
			FD/FD (prop.) & 33 & 128 & 96 & - & 6.0 & 0.3 & 22.1 \\
			Saving to \cite{Lowenborg_2006} & - & - & - & - & 91.9\% & 92.2\% & 84.8\% \\
			Saving to \cite{Moryakova_2023} & - & - & - & - & $-$17.1\% & 83.7\% & 7.1\% \\
			\hline
		\end{tabular}
		\vspace{-0.2cm}
		\label{tab:example_imp}
	\end{center}
\end{table}

\begin{table}[t]	
	\renewcommand{\arraystretch}{1.05}
	\caption{Design Complexity for the VBW Filter in the Example}
	\begin{center}
		\centering
		\begin{tabular}{l c c c c c c}
			\hline 
			Design/Impl. & $L$ & $N$ & $M$ & $R_{md}$ & $R_{ad}$ & Mem. \\
			\hline \hline
			TD/TD \cite{Lowenborg_2006} & 29 & - & - & 0 & 1 & 1 \\
			TD/FD(a) \cite{Moryakova_2023} & 29 & 128 & 100 & 5.2 & 5.1 & 640 \\	
			TD/FD(b) \cite{Moryakova_2023} & 29 & 128 & 100 & 7.7 & 15.4 & 75 \\	
			\hline
			FD/FD (prop.) & 33 & 128 & 0 & 0 & 0 & 15 \\
			\hline
		\end{tabular}
		\vspace{-0.2cm}
		\label{tab:example_des}
	\end{center}
\end{table}

\section{Conclusions}
\label{sec:conclusion}
This paper proposed an efficient approach to design an FC-based VBW filter using the HFSO. As shown through the design example, the proposed technique can significantly reduce both the implementation and design complexities by avoiding any arithmetic operations when the variable parameter $b$ is changed and by relaxing memory requirements.

\bibliographystyle{IEEEtran}
\bibliography{Reff}
\end{document}